\documentclass[twocolumn,aps,pra,showpacs,tightenlines]{revtex4-1}
\usepackage{amsmath}
\usepackage{amsfonts}
\usepackage{graphicx}
\usepackage{units}
\usepackage{color}
\usepackage[colorlinks,linkcolor=blue,anchorcolor=blue,citecolor=blue,urlcolor=blue]{hyperref}
\usepackage{multirow}
\begin{document}
\title{Optical normal-mode-induced phonon-sideband splitting in photon-blockade effect}

\author{Hong Deng}
\email{These authors contributed equally to this work.}
\affiliation{Key Laboratory of Low-Dimensional Quantum Structures and Quantum Control of
Ministry of Education, Key Laboratory for Matter Microstructure and Function of Hunan Province, Department of Physics and Synergetic Innovation Center for Quantum Effects and Applications, Hunan Normal University, Changsha 410081, China}

\author{Fen Zou}
\email{These authors contributed equally to this work.}
\affiliation{Key Laboratory of Low-Dimensional Quantum Structures and Quantum Control of
Ministry of Education, Key Laboratory for Matter Microstructure and Function of Hunan Province, Department of Physics and Synergetic Innovation Center for Quantum Effects and Applications, Hunan Normal University, Changsha 410081, China}

\author{Jin-Feng Huang}
\affiliation{Key Laboratory of Low-Dimensional Quantum Structures and Quantum Control of
Ministry of Education, Key Laboratory for Matter Microstructure and Function of Hunan Province, Department of Physics and Synergetic Innovation Center for Quantum Effects and Applications, Hunan Normal University, Changsha 410081, China}

\author{Jie-Qiao Liao}
\email{Corresponding author, jqliao@hunnu.edu.cn}
\affiliation{Key Laboratory of Low-Dimensional Quantum Structures and Quantum Control of
Ministry of Education, Key Laboratory for Matter Microstructure and Function of Hunan Province, Department of Physics and Synergetic Innovation Center for Quantum Effects and Applications, Hunan Normal University, Changsha 410081, China}

\date{\today}

\begin{abstract}
We study the photon-blockade effect in a loop-coupled optomechanical system consisting of two cavity modes and one mechanical mode. Here, the mechanical mode is optomechanically coupled to the two cavity modes, which are coupled with each other via a photon-hopping interaction. By treating the photon-hopping interaction as a perturbation, we obtain the analytical results of the eigenvalues and eigenstates of the system in the subspaces associated with zero, one, and two photons. We find a phenomenon of optical normal-mode-induced phonon-sideband splitting in the photon-blockade effect by analytically and numerically calculating the second-order correlation functions of the two cavity modes. This work not only presents a method to choose optimal driving frequency for photon blockade by tuning the photon-hopping interaction, but also provides a means to characterize the normal-mode splitting with cavity photon statistics.
\end{abstract}
\maketitle

\section{Introduction}

Single-photon sources are essential components for many rising quantum technologies such as quantum computation~\cite{knill2001Scheme}, quantum communication~\cite{kimble2008Quantum,sangouard2011Quantum}, and quantum cryptography~\cite{scarani2009Security}. The photon-blockade effect~\cite{imamoglu1997Strongly,carusotto2013Quantum}, the capture of a single photon in a system with strong optical nonlinearity hinders the injection of the second and subsequent photons, provides a way to realize single-photon sources. The key requirement for realizing photon blockade is that the optical nonlinearity should be stronger than the decay of the system. This is because the energy-level nonharmonicity should be resolved by the injected photons. In experiments, photon blockade can be characterized by the equal-time second-order correlation function $g^{(2)}(0)$. In quantum optics, the relation $g^{(2)}(0)<1$ [$g^{(2)}(0)>1$] is referred to as sub-Poissonian (super-Poissonian) statistics, and the relation $g^{(2)}(0)=1$ corresponds to the Poisson statistics~\cite{scully_zubairy_1997}. The correlation function $g^{(2)}(0)\ll1$ is considered as a signature of single-photon-blockade effect.

In the past two decades, photon blockade has been theoretically predicted in diverse nonlinear optical systems, e.g., cavity quantum electrodynamics (QED) systems~\cite{tian1992Quantum,birnbaum2005Photon,faraon2008Coherent,faraon2010Generation,ridolfo2012Photon,reinhard2012Strongly,peyronel2012Quantum,muller2015Coherent,
radulaski2017Photon,han2018Electromagnetic,zou2020Multiphoton}, circuit-QED systems~\cite{hoffman2011Dispersive,lang2011Observation,liu2014Blockade}, Kerr-type nonlinear cavities~\cite{liao2010Correlated,ghosh2019Dynamical}, and optomechanical systems~\cite{rabl2011Photon,liao2013Correlated,liao2013Photon,wang2015Tunable,zhu2018Controllable,zou2019Enhancement}. Particularly, photon blockade has been experimentally demonstrated in cavity- and circuit-QED systems, e.g., an optical cavity coupled to a trapped atom~\cite{birnbaum2005Photon}, a photonic crystal cavity coupled to a quantum dot~\cite{faraon2008Coherent,reinhard2012Strongly,muller2015Coherent}, and a microwave transmission-line resonator coupled to a superconducting artificial atom~\cite{hoffman2011Dispersive,lang2011Observation}. Recently, photon blockade has also been explored in coupled quantum optical systems, e.g., coupled Kerr-cavity systems~\cite{liew2010Single,bamba2011Origin,flayac2017Unconventional,zou2020Photon}, a bimodal cavity coupled to a quantum dot~\cite{zhang2014Optimal,liu2016Mode}, a two-level system coupled to two cavities~\cite{wang2017Phasemodulated}, and a linear cavity coupled to an optomechanical cavity~\cite{xu2013Antibunching,komar2013Singlephoton,li2019Nonreciprocal}. Nevertheless, the control of photon-blockade effect is a new research topics in this field.

In cavity optomechanical systems, it has been found that phonon sidebands can be used to adjust the photon-blockade effect~\cite{rabl2011Photon,liao2013Correlated}. To actively control the modulation in photon blockade, it is desired to find a tunable way to control the phonon sidebands in optomechanical systems~\cite{marquardt2007Quantum,
teufel2011Sideband,liao2012Spectrum,xiong2012Higher,liao2014Single}. To this end, in this paper we propose a normal-mode scheme~\cite{raizen1989Normal,thompson1992Observation,klinner2006Normal,
dobrindt2008Parametric,huang2009Normal,huang2010Normal,rossi2018Normal} to tune the resonance of phonon sidebands. Concretely, we study the photon blockade effect in a loop-coupled optomechanical system, which is composed of two cavity modes and one mechanical mode. Here the two cavity modes are linearly coupled to each other and the mechanical mode is coupled to each cavity mode via the radiation-pressure interaction. By analyzing the energy spectrum of the system, we find that the photon-hopping interaction will induce phonon-sideband splitting. We study the phenomenon of normal-mode induced phonon-sideband splitting by analytically and numerically calculating the equal-time second-order correlation functions of the two cavity modes. Specifically, we find that the optimal driving frequency for the photon blockade can be selected by adjusting the strength of the photon-hopping interaction between the two cavity modes~\cite{Chen2017Exceptional}. We also reveal a new method to characterize the normal-mode splitting phenomenon through cavity-field statistics.

The rest of this paper is organized as follows. In Sec.~\ref{modelsec}, we introduce the physical model and present the Hamiltonian. In Sec.~\ref{eigensystemsec}, we calculate the eigensystem of the Hamiltonian in the absence of driving and analyze the corresponding energy spectrum. In Sec.~\ref{pbsec}, we study photon blockade effect in this system by analytically and numerically calculating the equal-time second-order correlation functions of the two cavity modes. In Sec.~\ref{Discussion}, we give some discussions on the experimental implementation of the loop-coupled optomechanical system. Finally, we conclude this work in Sec.~\ref{conclusion}. We also present Appendixes~\ref{appendixa} and~\ref{appendixb} to show the detailed calculations of the eigensystem in the few-photon subspaces and the detailed derivation of the probability amplitudes, respectively.
\begin{figure}[tbp]
\center
\includegraphics[width=0.47 \textwidth]{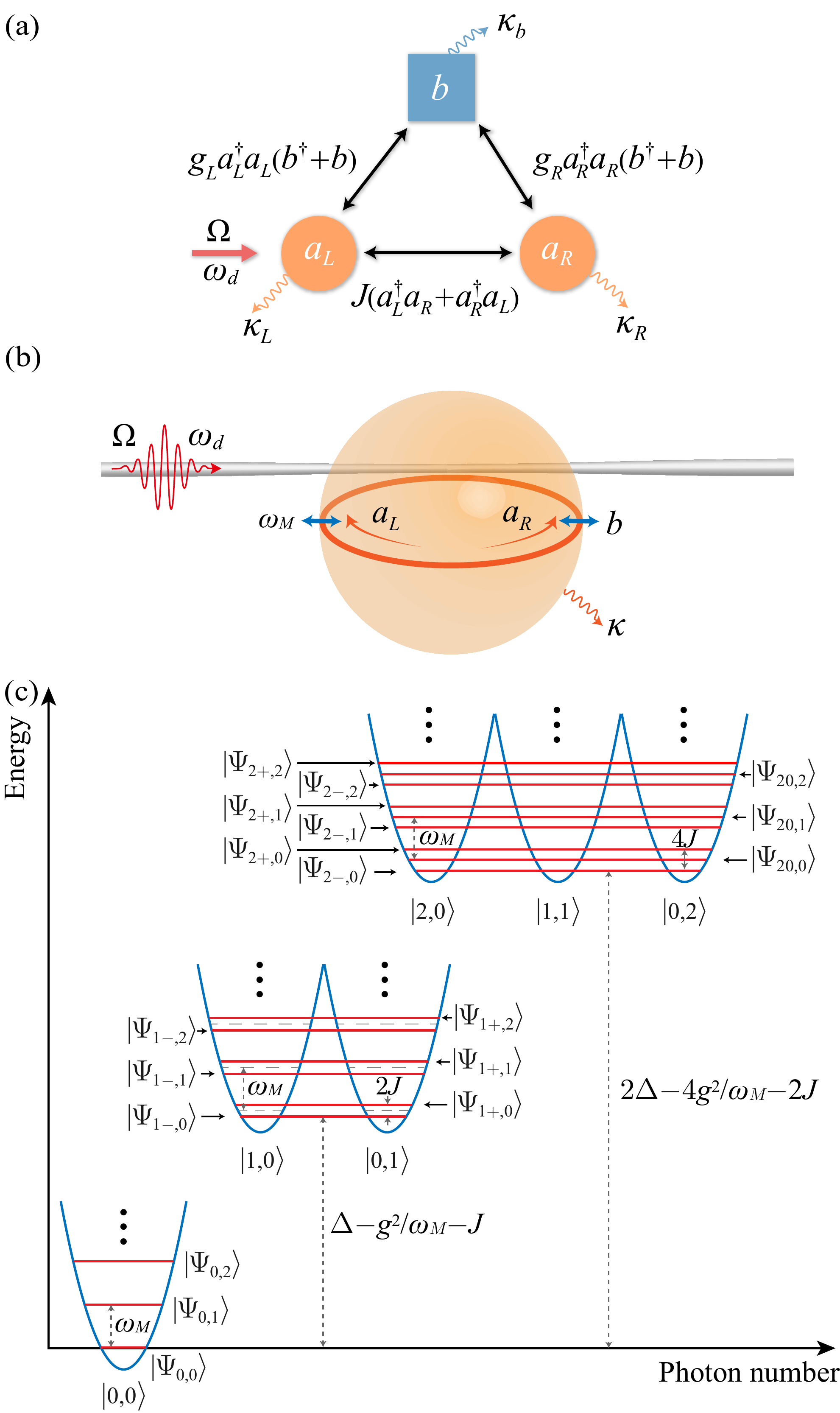}
\caption{(a) Cartoon diagram of the loop-coupled optomechanical system, which is composed of two cavity modes and one mechanical mode. The two circles and a square represent the degrees of freedom of the two cavity modes and the single mechanical mode, respectively. (b) Schematic diagram of the microsphere optomechanical cavity system, which is composed of two cavity modes (clockwise and counter-clockwise modes denoted by left- and right-rotation arrows, respectively) and one redial breathing mode. Here, the left-rotation cavity mode is driven by a monochromatic field through the side-coupled waveguide. (c) The eigenenergy spectrum of the Hamiltonian $\hat{H}_{\text{sys}}$ in these subspaces associated with zero, one, and two photons when $g_{L}=g_{R}=g$ and $\Delta_{L}=\Delta_{R}=\Delta$.}
\label{Fig1}
\end{figure}

\section{Model and Hamiltonian \label{modelsec}}

We consider a loop-coupled optomechanical system consisting of two cavity modes and one mechanical mode, as shown in Fig.~\ref{Fig1}(a). Here, we use two circles and a square to represent the degrees of freedom of the two cavity modes and the single mechanical mode, respectively. In particular, for notation convenience we adopt the subscripts ``$L$" and ``$R$" to mark the two cavity modes, which could be either the left- and right-hand rotation cavity modes (i.e., clockwise and counter-clockwise modes) in a microsphere optomechanical cavity~\cite{shen2018Reconfigurable}, or the cavity modes in the left and right subcavities in a ``membrane-in-the-middle" configuration optomechanical cavity~\cite{lee2015Multimode}. The two cavity modes are coupled to each other via a photon-hopping interaction and each cavity mode is coupled to the mechanical mode via the radiation-pressure interaction. When the cavity mode $a_{L}$ is coherently driven by a monochromatic field, the Hamiltonian (with $\hbar=1$) of the driven system reads~\cite{xu2015Optical}
\begin{eqnarray}
\hat{H} &=&\omega_{L}\hat{a}_{L}^{\dagger}\hat{a}_{L}+\omega_{R}\hat{a}_{R}^{\dagger}\hat{a}_{R}+\omega_{M}\hat{b}^{\dagger}\hat{b}
+J(\hat{a}_{L}^{\dagger}\hat{a}_{R}+\hat{a}_{R}^{\dagger}\hat{a}_{L}) \nonumber \\
&&-g_{L}\hat{a}_{L}^{\dagger}\hat{a}_{L}(\hat{b}^{\dagger}+\hat{b})-g_{R}\hat{a}_{R}^{\dagger}\hat{a}_{R}
(\hat{b}^{\dagger}+\hat{b})\nonumber \\
&&+\Omega(\hat{a}_{L}^{\dagger}e^{-i\omega_{d}t}+\hat{a}_{L}e^{i\omega_{d}t}), \label{H}
\end{eqnarray}
where $\hat{a}_{L}$ $(\hat{a}_{L}^{\dagger})$ and $\hat{a}_{R}$ $(\hat{a}_{R}^{\dagger})$ are the annihilation (creation) operators of the two cavity modes, with the corresponding resonance frequencies $\omega_{L}$ and $\omega_{R}$. The operator $\hat{b}$ $(\hat{b}^{\dagger})$ is the annihilation (creation) operator of the mechanical mode with the resonance frequency $\omega_{M}$. The parameter $J$ denotes the strength of the photon-hopping interaction between the two cavity modes. In typical experimental systems, it is difficult to tune the photon-hopping coupling strength once the device is fabricated. However, for the microtoroid optomechanical cavity system, the photon-hopping coupling strength can be tuned by controlling the scatterer (nano-tip) coupled to the microtoroid cavity~\cite{Chen2017Exceptional}. The parameter $g_{L}$ ($g_{R}$) describes the single-photon optomechanical-coupling strength between the cavity mode $a_{L}$ ($a_{R}$) and the mechanical mode. The last term in Eq.~(\ref{H}) describes monochromatic driving of the cavity mode $a_{L}$, with $\Omega$ and $\omega_{d}$ being the driving amplitude and driving frequency, respectively.

As we mentioned above, the present loop-coupled optomechanical model can be implemented in either a microsphere optomechanical cavity~\cite{shen2018Reconfigurable} or a ``membrane-in-the-middle" configuration optomechanical cavity~\cite{lee2015Multimode}. In the microsphere optomechanical cavity [Fig.~\ref{Fig1}(b)], the two cavity modes $a_{L}$ and $a_{R}$ correspond to the left-hand rotation (clockwise) mode and the right-hand rotation (counter-clockwise) mode, respectively. The mechanical mode is the radial breathing mode of the microsphere. The two cavity modes are coupled to the mechanical mode via the optomechanical interactions, and the coupling between the two cavity modes is induced by the optical backscattering mechanism~\cite{gorodetsky2000Rayleigh,kippenberg2002Modal}. The driving of the cavity mode $a_{L}$ can be applied through a waveguide side coupled to the microsphere cavity. On the other hand, for the ``membrane-in-the-middle" configuration optomechanical cavity~\cite{lee2015Multimode}, the two cavity modes $a_{L}$ and $a_{R}$ correspond to each cavity mode in the left and right subcavities, and the mechanical mode is realized by the oscillating membrane. The coupling between the two cavity modes is induced by the photon-tunneling interaction through the membrane, and the coupling between the cavity mode and the mechanical mode is realized by the optomechanical coupling. The driving to the cavity mode $a_{L}$ can be implemented by driving the left subcavity.

It should be emphasized that, through the loop-coupled optomechanical model can be implemented with either the microsphere optomechanical system or the ``membrane-in-the-middle" configuration optomechanical system, the relation between the optomechanical coupling strengths $g_{L}$ and $g_{R}$ in these two cases is different. For the microsphere optomechanical system, the optomechanical coupling strengths $g_{L}$ and $g_{R}$ take the same sign. Differently, for the ``membrane-in-the-middle" configuration optomechanical system, the coupling strengths $g_{L}$ and $g_{R}$ have opposite sign. This is because the vibration of the membrane increases (decreases) the length of one cavity and decreases (increases) the length of the other cavity at the same time. In the following discussions, we will mainly focus on the case of $g_{L}=g_{R}$, which is accessible for the microsphere optomechanical system. In addition, we present brief discussions on the case of $g_{L}=-g_{R}$ in Sec.~\ref{Discussion}. This case can be realized with the ``membrane-in-the-middle" optomechanical system. In these two cases, the main physical results concering the phonon-sideband splitting induced by optical normal modes can be exhibited.

In a frame rotating at the laser frequency $\omega_{d}$, the Hamiltonian of the system can be written as
\begin{equation}\label{HI}
\hat{H}_{I}=\hat{H}_{\text{sys}}+\Omega(\hat{a}_{L}^{\dagger}+\hat{a}_{L}),
\end{equation}
with
\begin{align}\label{Hsys}
\hat{H}_{\text{sys}}&=\Delta_{L}\hat{a}_{L}^{\dagger}\hat{a}_{L}+\Delta_{R}\hat{a}_{R}^{\dagger}\hat{a}_{R}+\omega_{M}\hat{b}^{\dagger}\hat{b}+J(\hat{a}
_{L}^{\dagger}\hat{a}_{R}+\hat{a}_{R}^{\dagger}\hat{a}_{L})   \nonumber \\
&\quad-g_{L}\hat{a}_{L}^{\dagger}\hat{a}_{L}(\hat{b}^{\dagger}+\hat{b})-g_{R}\hat{a}_{R}^{\dagger}\hat{a}_{R}(\hat{b}^{\dagger}+\hat{b}),
\end{align}
where $\Delta_{L}=\omega_{L}-\omega_{d}$ and $\Delta_{R}=\omega_{R}-\omega_{d}$ are the detunings of the two cavity-field frequencies with respect to the driving frequency. For Hamiltonian $\hat{H}_{\text{sys}}$, the total photon number operator $\hat{N}=\hat{a}_{L}^{\dagger}\hat{a}_{L}+\hat{a}_{R}^{\dagger}\hat{a}_{R}$ is a conserved quantity due to the commutative relation $[\hat{N},\hat{H}_{\text{sys}}]=0$. To study the photon-blockade effect in the loop-coupled optomechanical system, we only consider the weak-driving case. In this case, we can restrict the cavity modes within the low-excitation subspaces spanned by the basis states \{$\vert0,0\rangle_{LR}$\}, \{$\vert1,0\rangle_{LR}$, $\vert0,1\rangle_{LR}$\}, and \{$\vert 2,0\rangle_{LR}$, $\vert1,1\rangle_{LR}$, $\vert0,2\rangle_{LR}$\}. Here $\vert m,n\rangle_{LR}$ represents the state with $m$ photons in the cavity mode $a_{L}$ and $n$ photons in the cavity mode $a_{R}$. Moreover, in the absence of cavity-field driving and thermal excitation [it is reasonable to consider the vacuum environment for the cavity modes because $\hbar\omega_{L,R}/(k_{B}T)\gg1$], the two cavity modes will not be excited. Therefore, it is reasonable to assume that the two cavity modes are initially in the vacuum states. In this work, we consider the weak driving, then the average photon number in this system is much smaller than $1$, and then the system can be restricted into the few-photon subspaces.

\section{Eigensystem of the undriven system \label{eigensystemsec}}

In this section, we calculate the eigensystem of the Hamiltonian $\hat{H}_{\text{sys}}$ and analyze its energy spectrum. We also study the photon blockade effect because the conventional photon blockade effect is caused by the anharmonicity of the eigenenergy spectrum. For a general case, it is difficult to obtain the exact analytical results of the eigensystem of $\hat{H}_{\text{sys}}$. Here we will calculate the eigensystem of this Hamiltonian $\hat{H}_{\text{sys}}$ in the few-photon subspaces. In the absence of the photon-hopping term, the Hamiltonian of the system becomes
\begin{eqnarray}
\hat{H}_{\text{opt}}&=&\Delta_{L}\hat{a}_{L}^{\dagger}\hat{a}_{L}+\Delta_{R}\hat{a}_{R}^{\dagger}\hat{a}_{R}+\omega _{M}\hat{b}^{\dagger}\hat{b}\nonumber\\
&&-g_{L}\hat{a}_{L}^{\dagger}\hat{a}_{L}(\hat{b}^{\dagger}+\hat{b})
-g_{R}\hat{a}_{R}^{\dagger}\hat{a}_{R}(\hat{b}^{\dagger}+\hat{b}).
\end{eqnarray}
To diagonalize the Hamiltonian $\hat{H}_{\text{opt}}$, we introduce a conditional displacement operator $\hat{D}(\hat{\eta})=\exp[\hat{\eta}(\hat{b}^{\dagger}-\hat{b})]$, where the conditional displacement amplitude $\hat{\eta}$ is defined by
\begin{eqnarray}
\hat{\eta}&\equiv&(g_{L}\hat{a}_{L}^{\dagger}\hat{a}_{L}+g_{R}\hat{a}_{R}^{\dagger}\hat{a}_{R})/\omega_{M} \nonumber\\
&=&\sum_{m,n=0}^{\infty}\eta^{[m,n]}\vert m,n\rangle_{LR}\,_{LR}\langle m,n\vert,
\end{eqnarray}
with $\eta^{[m,n]}=(g_{L}m+g_{R}n)/\omega_{M}$.
Note that the mechanical displacement depends on the photon numbers in the two cavity modes.

The Hamiltonian $\hat{H}_{\text{opt}}$ can be diagonalized by a displacement transformation as
\begin{eqnarray}
\hat{\tilde{H}}_{\text{opt}}&=&\hat{D}^{\dagger}(\hat{\eta})\hat{H}_{\text{opt}}\hat{D}(\hat{\eta}) \nonumber\\
&=&\sum_{\mu=L,R}\Delta_{\mu}\hat{a}_{\mu}^{\dagger}\hat{a}_{\mu}+\omega_{M}\hat{b}^{\dagger}\hat{b}-\frac{(g_{L}\hat{a}_{L}^{\dagger}\hat{a}_{L}+g_{R}\hat{a}_{R}^{\dagger}\hat{a}_{R})^{2}}{\omega_{M}}.\nonumber\\
\end{eqnarray}
The eigensystem of $\hat{\tilde{H}}_{\text{opt}}$ can be obtained as $\hat{\tilde{H}}_{\text{opt}}\vert m,n\rangle_{LR}\vert k\rangle_{b}=E_{m,n,k}\vert m,n\rangle_{LR}\vert k\rangle_{b}$, with the corresponding eigenvalues
\begin{equation}\label{eigenvalues}
E_{m,n,k}=m\Delta_{L}+n\Delta_{R}+k\omega_{M}-(g_{L}m+g_{R}n)^{2}/\omega_{M}.
\end{equation}
Here $\vert k\rangle_{b}$ ($k=0,1,2,\cdots$) represent the number states of the mechanical mode.
Then the eigensystem of the Hamiltonian $\hat{H}_{\text{opt}}$ can be obtained as
\begin{equation}
\hat{H}_{\text{opt}}\vert m,n\rangle_{LR}\vert\tilde{k}(m,n)\rangle_{b}=E_{m,n,k}\vert m,n\rangle_{LR}\vert\tilde{k}(m,n)\rangle_{b}, \label{eigensystem}
\end{equation}
where we introduce the photon-number-dependent displaced phonon number states as
\begin{equation}\label{disphononnum}
\vert\tilde{k}(m,n)\rangle_{b}=\hat{D}(\eta^{[m,n]})\vert k\rangle_{b}.
\end{equation}
It can be proved that for a given two-mode photon number state $\vert m,n\rangle_{LR}$, the photon-number-dependent displaced phonon number states $\vert \tilde{k}(m,n)\rangle_{b}$ can constitute a complete orthonormal basis set in the Hilbert space of the mechanical mode, as shown by the two relations
$\sum_{k=0}^{\infty}\vert\tilde{k}(m,n)\rangle_{b}\!_{b}\langle\tilde{k}(m,n)\vert=I_{b}$ and $
_{b}\!\langle\tilde{k}(m,n)\vert\tilde{k}^{\prime}(m,n)\rangle\!_{b}=\delta_{k,k^{\prime}}$,
where $I_{b}$ is the identity operator of the mechanical mode. When $m\neq m^{\prime}$ or $n\neq n^{\prime}$, the inner product (the Franck-Condon factor) of these displaced number states for the mechanical mode can be calculated by
\begin{equation}
_{b}\!\langle\tilde{k}(m,n)\vert\tilde{k}^{\prime}(m^{\prime},n^{\prime})\rangle\!_{b}=\;\!_{b}\!\langle k\vert \hat{D}(\eta^{[m^{\prime},n^{\prime}]}-\eta^{[m,n]})\vert k^{\prime}\rangle\!_{b}.\label{matrixele}
\end{equation}
Here, the matrix elements in Eq.~(\ref{matrixele}) can be calculated based on the relation~\cite{deoliveira1990Properties}
\begin{align}
\label{displaopr}
\;_{b}\langle k\vert \hat{D}(\beta)\vert l\rangle_{b}=\left\{\begin{aligned}
\sqrt{\frac{k!}{l!}}e^{-\frac{\vert \beta\vert^{2}}{2}}(-\beta^{\ast})^{l-k}L_{k}^{l-k}(\vert \beta\vert^{2}),\hspace{0.1 cm}l\geq k, \\
\sqrt{\frac{l!}{k!}}e^{-\frac{\vert \beta\vert^{2}}{2}}(\beta)^{k-l}L_{l}^{k-l}(\vert \beta\vert^{2}),\hspace{0.1 cm} k>l,
\end{aligned}\right.
\end{align}
where $\hat{D}(\beta)=\exp(\beta \hat{b}^\dagger-\beta^{\ast}\hat{b})$ is a displacement operator and $L_{k}^{l}(\beta)$ are the associated Laguerre polynomials.

In the following we calculate the eigensystem of $\hat{H}_{\text{sys}}$ by adding the photon-tunneling term to $\hat{H}_{\text{opt}}$. In the eigenstate representation of $\hat{H}_{\text{opt}}$, the matrix element of $\hat{H}_{\text{sys}}$ can be obtained as
\begin{eqnarray}
&&_{b}\langle\tilde{k}(m,n)\vert_{LR}\!\langle m,n\vert\hat{H}_{\text{sys}}\vert m^{\prime},n^{\prime}\rangle_{LR}\vert\tilde{k}^{\prime}(m^{\prime},n^{\prime})\rangle\!_{b}\nonumber\\
&=&E_{m,n,k}\delta_{m,m^{\prime}}\delta_{n,n^{\prime}}\delta_{k,k^{\prime}}+J[\sqrt{(m^{\prime}+1)n^{\prime}}\delta_{m,m^{\prime}+1}\delta_{n,n^{\prime}-1}\nonumber\\
&&+\sqrt{m^{\prime}(n^{\prime}+1)}\delta_{m,m^{\prime}-1}\delta_{n,n^{\prime}+1}]\;_{b}\!\langle \tilde{k}(m,n)\vert \tilde{k}^{\prime}(m^{\prime},n^{\prime})\rangle_{b},\nonumber\\
\end{eqnarray}
where the Franck-Condon factors can be calculated with Eq.~(\ref{matrixele}). We note that the displaced-oscillator-number-state representation has been used to diagonalize the Hamiltonian of a coupled qubit-oscillator system~\cite{irish2005Dynamics}. In the case of $g_{L}\neq g_{R}$ and $\eta^{[m,n]}\ll1$, these matrix elements of $\hat{H}_{\text{sys}}$ can be solved approximately by using the zero-order approximation of $\eta^{[m,n]}$, i.e., $_{b}\!\langle \tilde{k}(m,n)\vert \tilde{k}^{\prime}(m^{\prime},n^{\prime})\rangle_{b}\approx\delta_{k,k^{\prime}}$. This approximation simplifies the calculation of the eigensystem of $\hat{H}_{\text{sys}}$ and it indicates that the photon tunneling only induces transitions with the same phonon-sideband index. The detailed calculations concerning few (zero, one, and two) photons will be given in Appendix~\ref{appendixa}. We point out that $_{b}\!\langle \tilde{k}(m,n)\vert \tilde{k}^{\prime}(m^{\prime},n^{\prime})\rangle_{b}=\delta_{m+n,m'+n'}\delta_{k,k^{\prime}}$ in the case of $g_{L}=g_{R}$.

In our following discussions, we consider the case of $g_{L}=g_{R}=g$ and $\Delta_{L}=\Delta_{R}=\Delta$. In this case, the  eigenvalues of $\hat{H}_{\text{sys}}$ can be obtained as
\begin{align}\label{varepsi}
\varepsilon_{0,k}&=k\omega_{M}, \nonumber\\
\varepsilon_{1\pm,k}&=\Delta+k\omega_{M}-\frac{g^{2}}{\omega_{M}}\pm J, \nonumber\\
\varepsilon_{20,k} &=2\Delta+k\omega_{M}-\frac{4g^{2}}{\omega_{M}}, \nonumber\\
\varepsilon_{2\pm,k}&=2\Delta+k\omega_{M}-\frac{4g^{2}}{\omega_{M}}\pm 2J.
\end{align}
The corresponding eigenstates are given by
\begin{align}
\vert\Psi_{0,k}\rangle&=\vert0,0\rangle_{LR}\vert k\rangle_{b}, \nonumber\\
\vert\Psi_{1\pm,k}\rangle&=\frac{1}{\sqrt{2}}(\vert1,0\rangle_{LR}\vert \tilde{k}(1,0)\rangle_{b}\pm\vert 0,1\rangle_{LR}\vert \tilde{k}(0,1)\rangle_{b}),\nonumber\\
\vert\Psi_{20,k}\rangle&=\frac{1}{\sqrt{2}}(\vert2,0\rangle_{LR}\vert\tilde{k}(2,0)\rangle_{b}-\vert 0,2\rangle_{LR}\vert \tilde{k}(0,2)\rangle_{b}), \nonumber\\
\vert\Psi_{2\pm,k}\rangle&=\frac{1}{2}(\vert2,0\rangle_{LR}\vert\tilde{k}(2,0)\rangle_{b}+\vert 0,2\rangle_{LR}\vert \tilde{k}(0,2)\rangle_{b})\nonumber\\
&\quad\pm\frac{1}{\sqrt{2}}\vert 1,1\rangle_{LR}\vert\tilde{k}(1,1)\rangle_{b}.
\end{align}
Equation~(\ref{varepsi}) indicates that the energy separation between two eigenvalues relating to neighboring phonon-sideband indexes $k$ and $k+1$ is $\omega_{M}$ in the same subspace (e.g., $\varepsilon_{1\pm,k+1}-\varepsilon_{1\pm,k}=\omega_{M}$), and the energy separation between two neighboring eigenvalues for the same phonon sideband is $2J$ in the single- and two-photon subspaces, i.e., $\varepsilon_{1+,k}-\varepsilon_{1-,k}=2J$, $\varepsilon_{2+,k}-\varepsilon_{20,k}=2J$, and $\varepsilon_{20,k}-\varepsilon_{2-,k}=2J$. Figure~\ref{Fig1}(c) shows the energy spectrum of the Hamiltonian $\hat{H}_{\text{sys}}$ in these subspaces associated with zero, one, and two photons. Owing to the anharmonicity of the energy spectrum in this system, the photon blockade effect can take place when the cavity is driven weakly. In addition, we can see from Fig.~\ref{Fig1}(c) that the photon-hopping interaction will induce a splitting between these phonon sidebands with the same index.

\section{Photon blockade effect \label{pbsec}}

In this section, we study the photon blockade effect in the two cavity modes by analytically and numerically calculating the equal-time second-order correlation function.

\subsection{Analytical results}
To include the influence of the dissipation of the two cavity modes on the photon blockade effect, we phenomenologically add the dissipation terms into Hamiltonian~(\ref{HI}). Then the effective Hamiltonian can be written as
\begin{equation}\label{Heff}
\hat{H}_{\text{eff}}=\hat{H}_{I}-i\frac{\kappa_{L}}{2}\hat{a}_{L}^{\dagger}\hat{a}_{L}-i\frac{\kappa_{R}}{2}\hat{a}_{R}^{\dagger}\hat{a}_{R},
\end{equation}
where $\kappa_{L}$ and $\kappa_{R}$ are the decay rates of the cavity modes $a_{L}$ and $a_{R}$, respectively. Here we only consider the dissipations of the two cavity modes and neglect the dissipation of the mechanical mode because the decay rates of two cavity modes are far larger than the mechanical dissipation. However, the dissipation of the mechanical mode will be included in the numerical results.

In the weak-driving regime ($\Omega\ll\kappa_{L}$), we can restrict the cavity modes within the low-excitation subspaces spanned by the basis states \{$\vert0,0\rangle_{LR}$\}, \{$\vert1,0\rangle_{LR}$, $\vert0,1\rangle_{LR}$\}, and \{$\vert 2,0\rangle_{LR}$, $\vert1,1\rangle_{LR}$, $\vert0,2\rangle_{LR}$\}. In this model, there are three typical sets of complete bases in the few-photon subspaces: (i) the first set is composed of bare states, i.e., $\vert m,n\rangle_{LR}\vert k\rangle_{b}$; (ii) the second set is composed of the eigenstates of Hamiltonian $H_{\text{opt}}$, i.e., $\vert m,n\rangle_{LR}\vert \tilde{k}(m,n)\rangle_{b}$; (iii) the third set is composed of the eigenstates of Hamiltonian $H_{\text{sys}}$, i.e., $\vert\Psi_{0,k}\rangle$, $\vert\Psi_{1\pm,k}\rangle$, $\vert\Psi_{20,k}\rangle$, and $\vert\Psi_{2\pm,k}\rangle$. For convenience, we choose the second basis set as basis states. In few-photon subspaces, a general state of the system can be expressed as
\begin{eqnarray}\label{psit}
\vert \psi(t)\rangle&=&\sum_{N=0}^{2}\sum_{m=0}^{N}\sum_{k=0}^{\infty}C_{m,N-m,k}(t)\vert m,N-m\rangle_{LR}  \nonumber\\
&&\times\vert \tilde{k}(m,N-m)\rangle_{b},
\end{eqnarray}
where the coefficients $C_{m,N-m,k}(t)$ are the probability amplitudes corresponding to states $\vert m,N-m\rangle_{LR}\vert \tilde{k}(m,N-m)\rangle$. Based on the Schr\"{o}dinger equation $i\vert\dot{\psi}(t)\rangle=\hat{H}_{\text{eff}}\vert\psi(t)\rangle$, the equations of motion of these probability amplitudes $C_{m,N-m,k}(t)$ can be obtained. Using the perturbation method~\cite{zou2020Multiphoton,zou2019Enhancement,zou2020Photon}, we can obtain the steady-state solutions of these probability amplitudes (see Appendix~\ref{appendixb}). Then the equal-time second-order correlation functions of the two cavity modes can be expressed as
\begin{eqnarray}\label{corrfuns}
g_{L}^{(2)}(0)\equiv\frac{\langle \hat{a}_{L}^{\dagger2}\hat{a}_{L}^{2} \rangle}{\langle \hat{a}_{L}^{\dagger}\hat{a}_{L}\rangle^2}=\frac{2P_{L,2}}{(P_{L,1}+2P_{L,2})^{2}}\approx\frac{2P_{L,2}}{P_{L,1}^2}, \nonumber\\
g_{R}^{(2)}(0)\equiv\frac{\langle \hat{a}_{R}^{\dagger2}\hat{a}_{R}^{2} \rangle}{\langle \hat{a}_{R}^{\dagger}\hat{a}_{R}\rangle^2}=\frac{2P_{R,2}}{(P_{R,1}+2P_{R,2})^{2}}\approx\frac{2P_{R,2}}{P_{R,1}^2},
\end{eqnarray}
where the state occupations of the two cavity modes are given by
\begin{eqnarray}\label{photoccups}
P_{L,1} \approx \sum_{k=0}^{\infty}\vert C_{1,0,k}\vert^{2}/\mathcal{N}, \quad
P_{R,1} \approx \sum_{k=0}^{\infty}\vert C_{0,1,k}\vert^{2}/\mathcal{N}, \quad \nonumber\\
P_{L,2} = \sum_{k=0}^{\infty}\vert C_{2,0,k}\vert^{2}/\mathcal{N}, \quad
P_{R,2} = \sum_{k=0}^{\infty}\vert C_{0,2,k}\vert^{2}/\mathcal{N}, \quad
\end{eqnarray}
with the normalization constant
\begin{eqnarray}
\mathcal{N}&=&\sum_{N=0}^{2}\sum_{m=0}^{N}\sum_{k=0}^{\infty}\vert C_{m,N-m,k}\vert^{2}.
\end{eqnarray}
In the weak-driving case, the normalization constant $\mathcal{N}\approx1$. Inserting Eq.~(\ref{proampper}) and Eq.~(\ref{photoccups}) into Eq.~(\ref{corrfuns}), we find the analytical expressions of the correlation functions for the two cavity modes are independent of the driving strength. Therefore, the phenomenon of phonon-sideband splitting in photon blockade is a general physical effect in the weak driving case.
\begin{figure*}[htbp]
\center
\includegraphics[width=14cm]{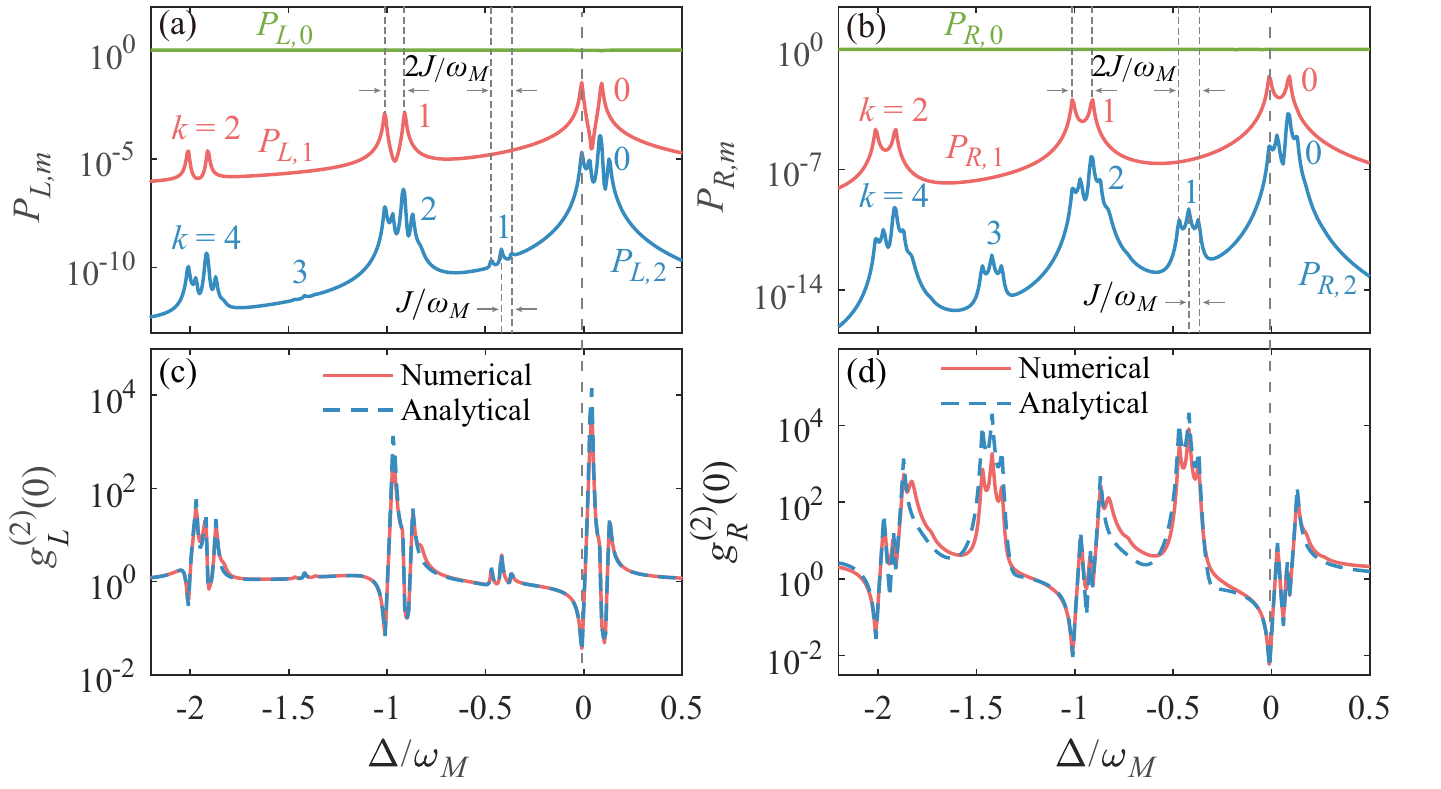}
\caption{The state occupations (a) $P_{L,m=0,1,2}$ and (b) $P_{R,m=0,1,2}$ as functions of the driving detuning $\Delta/\omega_{M}$. The correlation functions (c) $g^{(2)}_{L}(0)$ and (d) $g^{(2)}_{R}(0)$ as functions of the driving detuning $\Delta/\omega_{M}$. Blue dashed curves and red solid curves
are plotted based on the analytical and numerical results, respectively. The gray dashed lines correspond to the position of $\Delta=g^{2}/\omega_{M}-J$. Other parameters used are $\Delta_{L}=\Delta_{R}=\Delta$, $g_{L}/\omega_{M}=g_{R}/\omega_{M}=0.2$, $J/\omega_{M}=0.05$, $\kappa_{L}/\omega_{M}=\kappa_{R}/\omega_{M}=0.01$, $\kappa_{b}/\omega_{M}=0.001$, $\bar{n}_{b}=0$, and $\Omega/\kappa_{L}=0.2$.}
\label{Fig2}
\end{figure*}

\subsection{Numerical results}

In the above analytical calculations, the jumping terms are neglected with the effective Hamiltonian method, and the dissipation of the mechanical mode is neglected. Below, we include the two neglected elements by numerically solving the quantum master equation. Concretely, we assume that the two cavity modes are connected with two individual vacuum baths and the mechanical mode is connected with a heat bath at temperature $T$. Then the dynamics of the system is governed by the quantum master equation
\begin{align}
\dot{\hat{\rho}}&=i[\hat{\rho},\hat{H}_{I}]+\frac{\kappa_{L}}{2}\mathcal{L}_{\hat{a}_{L}}[\hat{\rho}]
+\frac{\kappa_{R}}{2}\mathcal{L}_{\hat{a}_{R}}[\hat{\rho}] \nonumber\\ &\quad+\frac{\kappa_{b}}{2}(\bar{n}_{b}+1)\mathcal{L}_{\hat{b}}[\hat{\rho}]
+\frac{\kappa_{b}}{2}\bar{n}_{b}\mathcal{L}_{\hat{b}^{\dagger}}[\hat{\rho}],\label{MEQ}
\end{align}
where $\hat{H}_{I}$ is the Hamiltonian given in Eq.~(\ref{HI}), $\kappa_{L}$ ($\kappa_{R}$) is the decay rate of the cavity mode $a_{L}$ ($a_{R}$), and $\kappa_{b}$ is the dissipation rate of the mechanical mode. The parameter $\bar{n}_{b}$ is the average thermal phonon number associated with the mechanical mode given by $\bar{n}_{b}=[\exp(\hbar\omega_{M}/k_{B}T)-1]^{-1}$, where $T$ is the temperature of the heat bath and $k_{B}$ is the Boltzmann constant.  $\mathcal{L}_{\hat{o}}[\hat{\rho}]=(2\hat{o}\hat{\rho}\hat{o}^{\dagger}-\hat{o}^{\dagger }\hat{o}\hat{\rho}-\hat{\rho}\hat{o}^{\dagger}\hat{o})$ are the Lindblad superoperators with $\hat{o}=\hat{a}_{L}$, $\hat{a}_{R}$, $\hat{b}$, and $\hat{b}^{\dagger}$. The Lindblad superoperators $\mathcal{L}_{\hat{a}_{L}}[\hat{\rho}]$, $\mathcal{L}_{\hat{a}_{R}}[\hat{\rho}]$, and $\mathcal{L}_{\hat{b}}[\hat{\rho}]$ describe the losses of the two cavity modes and the mechanical mode, while $\mathcal{L}_{\hat{b}^{\dagger}}[\hat{\rho}]$ describes the mechanical thermal excitation.

By numerically solving the quantum master equation~(\ref{MEQ}) with the Python package QuTiP~\cite{johansson2012QuTiP,johansson2013QuTiP}, we can get the steady-state density operator $\hat{\rho}_{\text{ss}}$ of the system. Then the photon-number-state occupations of the two cavity modes can be obtained by $P_{\mu,m=0,1,2}=\text{Tr}[\vert m\rangle_{\mu}\!_{\mu}\langle m\vert\hat{\rho}_{\text{ss}}]$ $(\mu=L,R)$. Similarly, we can obtain the equal-time second-order correlation functions of the two cavity modes by $g_{\mu}^{(2)}(0)=\text{Tr}(\hat{a}^{\dagger}_{\mu}\hat{a}^{\dagger}_{\mu}\hat{a}_{\mu}\hat{a}_{\mu}\hat{\rho}_{\text{ss}})/
[\text{Tr}(\hat{a}^{\dagger}_{\mu}\hat{a}_{\mu}\hat{\rho}_{\text{ss}})]^2$ with $\mu=L,R$. To observe the photon blockade effect of this system, we assume that the two optomechanical coupling strengths of the system enters the single-photon strong-coupling regime. Although the single-photon strong-coupling regime has not been experimental realized, some methods have been proposed to amplify the optomechanical coupling strength induced by a single photon. These methods include the introduction of collective mechanical modes~\cite{Xuereb2012Strong}, the modulation of system parameters~\cite{Liao2014Modulated,Liao2015Enhancement}, the utilizing of the Josephson nonlinearity~\cite{Rimberg2014cavity,Rimberg2014Enhancing,Pirkkalainen2015Cavity}, the coupling coefficient amplification with either the squeezing~\cite{Lv2015Squeezed,Lemonde2016Enhanced,Li2016Enhanced} or displacement transformations~\cite{Liao2020Generalized}, and the utilizing of quantum feedback technique~\cite{Wang2017Enhancing}.

To study the photon-blockade effect in the two cavity modes, we plot the photon-number state occupations $P_{\mu,m=0,1,2}$ ($\mu=L,R$) of the two cavity modes as functions of the driving detuning $\Delta/\omega_{M}$ in Figs.~\ref{Fig2}(a) and~\ref{Fig2}(b). It can be seen that the photon-number-state occupations of the two cavity modes satisfy the relations $P_{\mu,0}\approx1$ and $P_{\mu,0}\gg P_{\mu,1}\gg P_{\mu,2}$ ($\mu=L,R$) in the weak-driving case. Furthermore, we observe that for each phonon sideband, there are two subpeaks in the curves of the single-photon state occupations $P_{\mu,1}$. Hereafter, we call the envelope of these subpeaks as a main peak associated with the corresponding phonon sideband. These subpeaks can be qualitatively understood from the energy-level diagram shown in Fig.~\ref{Fig1}(c). By analyzing Fig.~\ref{Fig1}(c), we find that these subpeaks in the curves of $P_{\mu,1}$ correspond to the single-photon resonance conditions $\Delta+k\omega_{M}-g^{2}/\omega_{M}\pm J=0$, where $k=0,1,2,\cdots$ is the phonon-sideband index, as marked in Figs.~\ref{Fig2}(a) and~\ref{Fig2}(b). The locations of these subpeaks are $\Delta=g^{2}/\omega_{M}\mp J-k\omega_{M}$, i.e., corresponding to the single-photon resonance transitions $\vert\Psi_{0,0}\rangle\rightarrow\vert\Psi_{1\pm,k}\rangle$.

For the two-photon state occupations $P_{\mu,2}$, we see that there are four (three) subpeaks in each even (odd) phonon sideband. In each odd phonon sideband, these three subpeaks correspond to the two-photon resonance conditions $2\Delta+k\omega_{M}-4g^{2}/\omega_{M}\pm 2J=0$ and $2\Delta+k\omega_{M}-4g^{2}/\omega_{M}=0$, respectively. Then the locations of these three peaks are given by $\Delta=2g^{2}/\omega_{M}\mp J-k\omega_{M}/2$ and $\Delta=2g^{2}/\omega_{M}-k\omega_{M}/2$, i.e., corresponding to the two-photon resonant transitions $\vert\Psi_{0,0}\rangle\rightarrow\vert\Psi_{2s,k}\rangle$ $(s=\pm,0)$. In each even phonon sideband, the locations of these three subpeaks are determined by the two-photon processes, the remaining subpeak is induced by the single-photon resonant transition $\vert\Psi_{0,0}\rangle\rightarrow\vert\Psi_{1+,k}\rangle$. Here only four subpeaks can be observed in each even photon sideband because the subpeaks corresponding to the single-photon transition processes $\vert\Psi_{0,0}\rangle\rightarrow\vert\Psi_{1-,k}\rangle$ and two-photon transition processes $\vert\Psi_{0,0}\rangle\rightarrow\vert\Psi_{20,2k}\rangle$ cannot be distinguished for the parameters used in our simulations.

Based on the above analyses, we know that the distances between two neighboring subpeaks in $P_{L(R),1}$ and $P_{L(R),2}$ associated with a certain $k$ are $2J$ and $J$, respectively. In addition, the distance between the main peaks relating to neighboring phonon sideband indexes $k$ and $k+1$ is $\omega_{M}$. Therefore, for resolving these subpeaks associated with the same phonon-sideband index (in the same main peak), the system should work in the regime of $J\gg\kappa$. However, for resolving the phonon sidebands relating to neighboring sideband indexes (namely resolving these main peaks), the resolved-sideband condition $\omega_{M}\gg \kappa$ should be satisfied. These relations can be obtained by analyzing the eigen-energy spectrum of the system, as shown in Fig.~\ref{Fig1}(c).
\begin{figure}[tbp]
\center
\includegraphics[width=8.5cm]{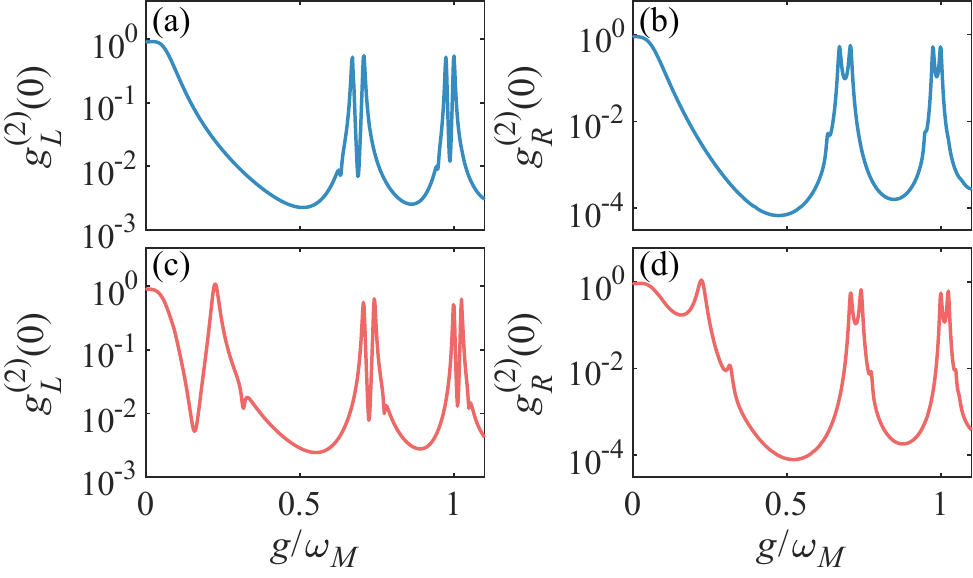}
\caption{The correlation functions $g^{(2)}_{L}(0)$ and $g^{(2)}_{R}(0)$ as functions of the optomechanical-coupling strength $g/\omega_{M}$ at (a),(b) $\Delta=g^{2}/\omega_{M}-J$ and (c),(d) $\Delta=g^{2}/\omega_{M}+J$. Other parameters used are $\Delta_{L}=\Delta_{R}=\Delta$, $g_{L}=g_{R}=g$, $J/\omega_{M}=0.05$, $\kappa_{L}/\omega_{M}=\kappa_{R}/\omega_{M}=0.01$, $\kappa_{b}/\omega_{M}=0.001$, $\bar{n}_{b}=0$, and $\Omega/\kappa_{L}=0.2$.}
\label{Fig3}
\end{figure}

To clarify the optimal driving frequency of the photon-blockade effect, in Figs.~\ref{Fig2}(c) and~\ref{Fig2}(d) we plot the equal-time second-order correlation functions $g_{\mu=L,R}^{(2)}(0)$ as functions of the driving detuning $\Delta/\omega_{M}$. Here, the blue dashed curves and red solid curves
are plotted based on the analytical and numerical results, respectively. It can be seen that the analytical result has an excellent agreement with the numerical result. We also observe a sequence of super-Poissonian $(g_{\mu}^{(2)}(0)>1)$ and sub-Poissonian $(g_{\mu}^{(2)}(0)<1)$ photon statistics. In addition, it can be found that the locations of these dips and peaks of the correlation functions $g_{\mu}^{(2)}(0)$ correspond to single- and two-photon resonance processes, respectively. In the single-photon resonance case, a single photon can be resonantly injected into the cavity, while the injection of the second photon is largely suppressed owing to the anharmonicity of the energy spectrum. This indicates that the photon-blockade effect ($g_{\mu}^{(2)}(0)\ll1$) occurs under the single-photon resonance condition. In particular, we find that the optimal driving frequencies of the photon-blockade effect can be selected by tuning the coupling strength between the two cavity modes. The reason is that the optimal driving frequencies ($\Delta=g^{2}/\omega_{M}\mp J-k\omega_{M}$) of the photon-blockade effect depend on the photon-hopping strength $J$. In Fig.~\ref{Fig2}, we use the gray dashed lines to mark the single-photon resonance process $\vert\Psi_{0,0}\rangle\rightarrow\vert\Psi_{1+,0}\rangle$, corresponding to the peak or dip located at $\Delta=g^{2}/\omega_{M}-J$.
\begin{figure}[tbp]
\center
\includegraphics[width=8.5cm]{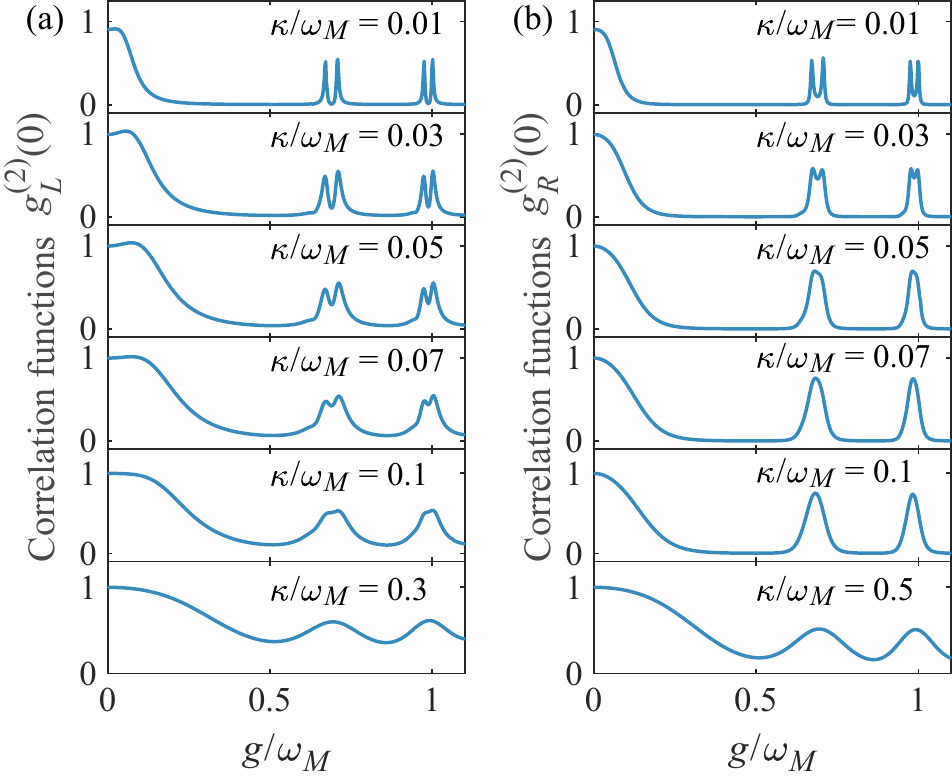}
\caption{The correlation functions (a) $g^{(2)}_{L}(0)$ and (b) $g^{(2)}_{R}(0)$ as functions of $g/\omega_{M}$ at various values of $\kappa/\omega_{M}$ when $\Delta=g^{2}/\omega_{M}-J$. Other parameters used are $\Delta_{L}=\Delta_{R}=\Delta$, $g_{L}=g_{R}=g$, $J/\omega_{M}=0.05$, $\kappa_{L}=\kappa_{R}=\kappa$, $\kappa_{b}/\omega_{M}=0.001$, $\bar{n}_{b}=0$, and $\Omega/\omega_{M}=0.002$.}
\label{Fig4}
\end{figure}

We proceed to study the influence of the optomechanical-coupling strength on the photon blockade effect. In Fig.~\ref{Fig3}, we plot the correlation functions $g_{\mu=L,R}^{(2)}(0)$ as functions of $g/\omega_{M}$ at the single-photon resonant transitions $\Delta=g^{2}/\omega_{M}\mp J$, i.e., $\vert\Psi_{0,0}\rangle\rightarrow\vert\Psi_{1\pm,0}\rangle$. The blue and red solid curves correspond to the cases of $\Delta=g^{2}/\omega_{M}-J$ and $\Delta=g^{2}/\omega_{M}+J$, respectively. It can be observed from Fig.~\ref{Fig3} that the values of $g_{\mu}^{(2)}(0)$ are approximately equal to 1 when $g/\omega_{M}\lesssim0.05$, which means that the photon-blockade effect cannot occur. In addition, there are several resonance peaks at specific values of $g/\omega_{M}$. Due to the modulation of the phonon sidebands, the single- and two-photon resonant transitions can be simultaneously induced. Hence, the locations of these resonance peaks in Figs.~\ref{Fig3}(a) and~\ref{Fig3}(b) [Figs.~\ref{Fig3}(c) and~\ref{Fig3}(d)] correspond to the two-photon resonant transitions $\vert\Psi_{1+,0}\rangle\rightarrow\vert\Psi_{2s,k}\rangle$ ($\vert\Psi_{1-,0}\rangle\rightarrow\vert\Psi_{2s,k}\rangle$). By analyzing the energy spectrum of the system, the optomechanical-coupling strengths corresponding to the single- and two-photon resonant transitions can be obtained as $g^{[k+]}=[(k\omega_{M}^{2}-4J\omega_{M})/2]^{1/2},[(k\omega_{M}^{2}-2J\omega_{M})/2]^{1/2}$, $(k\omega_{M}^{2}/2)^{1/2}$, and $g^{[k-]}=(k\omega_{M}^{2}/2)^{1/2},[(k\omega_{M}^{2}+2J\omega_{M})/2]^{1/2}$, $[(k\omega_{M}^{2}+4J\omega_{M})/2]^{1/2}$, respectively. Here, the parameter conditions of $g^{[k+]}$ ($g^{[k-]}$) are determined simultaneously by the single-photon resonant transition $\vert\Psi_{0,0}\rangle\rightarrow\vert\Psi_{1+,0}\rangle$ ($\vert\Psi_{0,0}\rangle\rightarrow\vert\Psi_{1-,0}\rangle$) and two-photon resonant transition $\vert\Psi_{1+,0}\rangle\rightarrow\vert\Psi_{2s,k}\rangle$ ($\vert\Psi_{1-,0}\rangle\rightarrow\vert\Psi_{2s,k}\rangle$). It follows from the relations $g^{[k+]}$ and $g^{[k-]}$ that the distance between the two subpeaks depends on the photon-hopping strength $J$ when the phonon-sideband index $k$ is given. This implies that the phenomenon of optical normal-mode-induced phonon-sideband splitting can be observed in the second-order correlation function.

We also analyze how the photon-blockade effect depends on the cavity-field decay rate. The correlation functions $g_{\mu=L,R}^{(2)}(0)$ are plotted in Fig.~\ref{Fig4} as functions of $g/\omega_{M}$ at various values of $\kappa/\omega_{M}$ when $\Delta=g^{2}/\omega_{M}-J$. We observe that due to the modulation of the phonon sidebands, the correlation functions $g_{\mu}^{(2)}(0)$ exhibit several resonance subpeaks at specific values of $g/\omega_{M}$. Since the values of the peaks in $g^{[k+]}=[(k\omega_{M}^{2}-4J\omega_{M})/2]^{1/2}$ are very small, only two subpeaks can be observed in each phonon sideband. Particularly, we find that the two subpeaks relating to each phonon sideband coalesce into a main peak with the increase of $\kappa/\omega_{M}$. When the decay rate is much larger than the photon-hopping strength $\kappa\gg J$, the normal mode becomes unresolved and hence the normal-mode splitting phenomenon disappears. In addition, we find that the photon blockade effect in the two cavity modes attenuates with the increase of $\kappa/\omega_{M}$. In particular, the correlation function of the cavity mode $a_{L}$ is more robust than that of the cavity mode $a_{R}$ against the cavity-field decay.
\begin{figure}[tbp]
\center
\includegraphics[width=7.5cm]{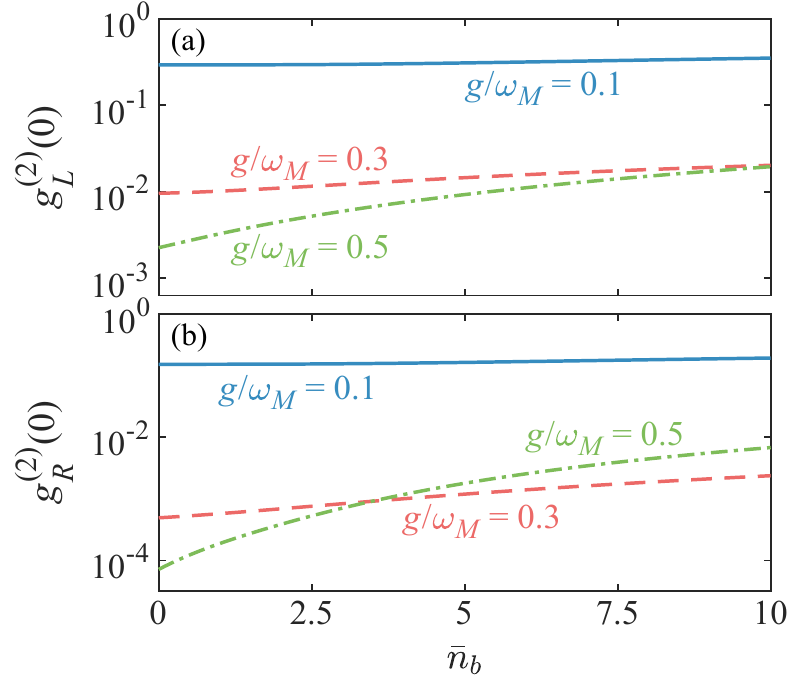}
\caption{The correlation functions (a) $g^{(2)}_{L}(0)$ and (b) $g^{(2)}_{R}(0)$ as functions of the thermal phonon number $\bar{n}_{b}$ at various values of $g/\omega_{M}$ when $\Delta=g^{2}/\omega_{M}-J$. Other parameters used are $\Delta_{L}=\Delta_{R}=\Delta$, $g_{L}=g_{R}=g$, $J/\omega_{M}=0.05$, $\kappa_{L}/\omega_{M}=\kappa_{R}/\omega_{M}=0.01$, $\kappa_{b}/\omega_{M}=0.001$, and $\Omega/\kappa_{L}=0.2$.}
\label{Fig5}
\end{figure}

Below, we study the influence of the mechanical thermal noise on the photon-blockade effect in the two cavity modes. In Fig.~\ref{Fig5}, we show the correlation functions $g_{\mu=L,R}^{(2)}(0)$ as functions of the thermal phonon number $\bar{n}_{b}$ at various values $g/\omega_{M}$ when $\Delta=g^{2}/\omega_{M}-J$. Here, the values of $g_{\mu}^{(2)}(0)$ increase slowly with the increase of $\bar{n}_{b}$. This indicates that the thermal noise will weaken the photon-blockade effect. However, the photon-blockade effect can be enhanced with the increase of $g/\omega_{M}$ when $\bar{n}_{b}=0$.

\section{Discussions \label{Discussion}}
In this section, we present some discussions on the experimental realization of the loop-coupled optomechanical system. Currently, the loop-coupled optomechanical system has been experimentally implemented in a microsphere optomechanical cavity system~\cite{shen2018Reconfigurable}. As shown in Ref.~\cite{shen2018Reconfigurable}, the microresonator supports a pair of degenerate clockwise and counter-clockwise traveling-wave whispering-galley modes. The radial breathing (mechanical) mode modulates the resonant frequencies of the clockwise and counter-clockwise modes by changing the circumference of the microsphere, and then the optomechanical interaction exists between the radial breathing mode and the clockwise (counter-clockwise) mode. In this system, the photon-hopping interaction between the clockwise and counter-clockwise modes is realized due to the optical backscattering~\cite{gorodetsky2000Rayleigh,kippenberg2002Modal}. In the microsphere system, the resonance frequency of the clockwise (counter-clockwise) mode is $\omega_{c}/2\pi=384.6$ THz and the decay rate is $\kappa/2\pi=16.2$ MHz. The coupling strength between the clockwise and counter-clockwise modes is $J/2\pi=1.5$ MHz. The radial breathing mode has a resonance frequency of $\omega_{M}/2\pi=90.47$ MHz and a dissipation rate of $\gamma/2\pi=22$ kHz. In particular, in the microsphere optomechanical cavity system, the two coupling strengths satisfy the relation $g_{L}=g_{R}$. In our simulations, we use the experimentally feasible parameters: $g_{L}/\omega_{M}=g_{R}/\omega_{M}=0.2$, $J/\omega_{M}=0.05$, $\kappa_{L}/\omega_{M}=\kappa_{R}/\omega_{M}=\kappa/\omega_{M}=0.01$, $\kappa_{b}/\omega_{M}=0.001$, and $\Omega/\kappa=0.2$. Here we point out that the two optomechanical coupling strengths of the system are assumed to reach the ultrastrong coupling regime for observing photon blockade effect. In realistic systems, the optomechanical coupling is much smaller than the decay rates of the cavity modes. In addition, the value of the ratio $J/\omega_{M}$ used in our simulations is of the same order of the real experimental parameter. These parameters should be accessible with the near future experimental conditions. Physically, despite the photon-hopping interaction strength induced by the optical backscattering being low, these subpeaks associated with the same phonon-sideband index can be resolved because the linewidth of the cavity is smaller than the subpeak splitting.
\begin{figure}[tbp]
\center
\includegraphics[width=7.5cm]{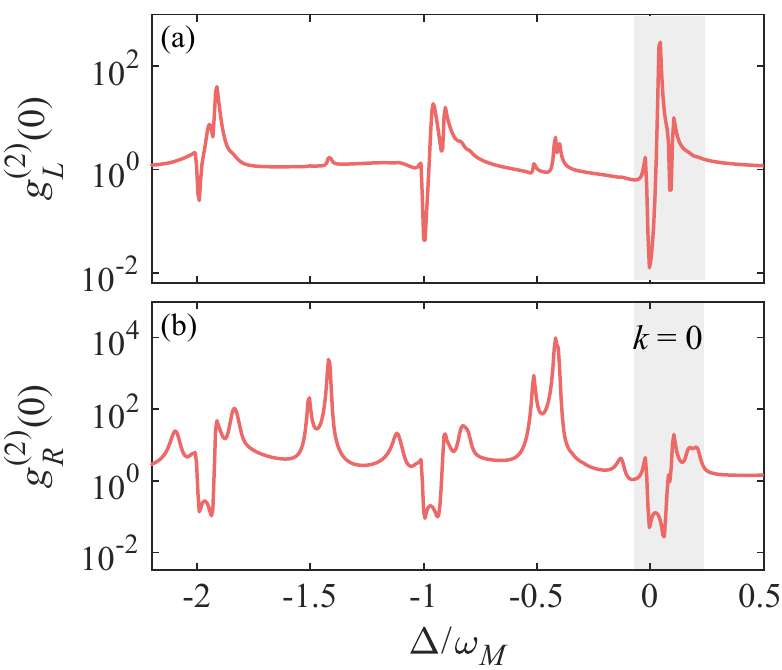}
\caption{The correlation functions (a) $g^{(2)}_{L}(0)$ and (b) $g^{(2)}_{R}(0)$ as functions of the driving detuning $\Delta/\omega_{M}$. The gray areas correspond to the results of $k=0$. Other parameters used are $\Delta_{L}=\Delta_{R}=\Delta$, $g_{L}/\omega_{M}=-g_{R}/\omega_{M}=-0.2$, $J/\omega_{M}=0.05$, $\kappa_{L}/\omega_{M}=\kappa_{R}/\omega_{M}=0.01$, $\kappa_{b}/\omega_{M}=0.001$, $\bar{n}_{b}=0$, and $\Omega/\kappa_{L}=0.2$.}
\label{Fig6}
\end{figure}

In addition, the loop-coupled optomechanical system has been experimentally implemented in a ``membrane-in-the-middle" configuration optomechanical cavity system~\cite{lee2015Multimode}. As shown in Ref.~\cite{lee2015Multimode}, there are two cavity modes (left and right subcavity modes) in the “membrane-in-the-middle” configuration optomechanical cavity. The left and right cavity modes are coupled to each other via a photon-hopping interaction. The vibration of the membrane can tune the resonant frequencies of the two cavity modes, and then the radiation-pressure interaction exists between the mechanical mode and the left (right) subcavity mode. In this system~\cite{lee2015Multimode}, the decay rates of the left and right subcavity modes are $\kappa_{L}/2\pi=1$ MHz and $\kappa_{R}/2\pi=1.3$ MHz, and the photon-hopping interaction strength between two cavity modes is $J/2\pi=4.6$ MHz. The mechanical mode has a resonance frequency of $\omega_{M}/2\pi=354.6$ kHz with a quality factor $Q=10^{5}$, and the two optomechanical coupling strengths satisfy the relation $g_{L}=-g_{R}$. Note that here we also assume that the optomechanical coupling is working in the single-photon strong coupling regime, which is still not accessible in realistic physical systems. To investigate the photon blockade effect of the left and right cavities in the case of $g_{L}=-g_{R}$, we plot the second-order correlation functions $g_{\mu=L,R}^{(2)}(0)$ as functions of the driving detuning $\Delta/\omega_{M}$ in Figs.~\ref{Fig6}(a) and~\ref{Fig6}(b). Similarly, one can see that a sequence of super-Poissonian $(g_{\mu}^{(2)}(0)>1)$ and sub-Poissonian $(g_{\mu}^{(2)}(0)<1)$ photon statistics. Compared with Figs.~\ref{Fig2}(c) and~\ref{Fig2}(d), we find that the difference is that the positions of the resonance peaks in $g_{\mu=L,R}^{(2)}(0)$ are moved due to the change of the eigenenergy spectrum in the case of $g_{L}=-g_{R}$. For the parameters used in our simulations, when the phonon sideband is not considered (i.e., $k=0$), the locations of the three main peaks [the two-photon resonant transitions $\vert\Psi_{0,0}\rangle\rightarrow\vert\Psi_{2s,0}\rangle$ $(s=\pm,0)$] are respectively $\Delta/\omega_{M}=-0.02403$, $0.08$, and $0.10403$, and the locations of the two dips (the single-photon resonant transitions $\vert\Psi_{0,0}\rangle\rightarrow\vert\Psi_{1\pm,0}\rangle$) are $\Delta/\omega_{M}=-0.01$ and $0.09$. In the single-photon resonance transition, we can see that photon blockade effect takes place in the two cavity modes.

\section{Conclusion \label{conclusion}}

In conclusion, we studied the photon-blockade effect in a loop-coupled optomechanical system, in which the two cavity modes are coupled to each other by a photon-hopping interaction and the mechanical mode is coupled to the two cavity modes through the radiation-pressure interaction. Here, the left cavity mode is weakly driven by a monochromatic laser field. We obtained the analytical result of the eigensystem in the weak photon-hopping case. By analyzing the energy spectrum of the system, we found that the photon-hopping interaction will induce normal-mode splitting in the subspaces associated with the same phonon-sideband index. In particular, we found an interesting phenomenon of optical normal-mode-induced phonon-sideband splitting in the second-order correlation function of the cavity modes. We also found that the photon-blockade effect in the two cavity modes can be observed in the single-photon resonant driving case. This scheme not only shows that the optimal driving frequency for photon blockade can be selected by tuning the coupling strength of the photon-hopping interaction, but also provides an experimental means to observe the normal-mode splitting effect through the correlation function of cavity fields.

\begin{acknowledgments}
J.-Q.L. is supported in part by National Natural Science Foundation of China (Grants No. 11822501, No. 11774087, No. 12175061, and No. 11935006), Hunan Science and Technology Plan Project (Grant No. 2017XK2018), and the Science and Technology Innovation Program of Hunan Province (Grant No. 2020RC4047). J.-F.H. is supported in part by the National Natural Science Foundation of China (Grant No.~12075083), Scientific Research Fund of Hunan Provincial Education Department (Grant No.~18A007), and Natural Science Foundation of Hunan Province, China (Grant No.~2020JJ5345).
\end{acknowledgments}

\appendix

\section{Eigensystem in the few-photon subspaces \label{appendixa}}
In this Appendix, we present the eigensystem of the Hamiltonian $\hat{H}_{\text{sys}}$ in the few-photon subspaces.

\subsection{Eigensystem in the zero- and single-photon subspaces \label{eigen01photon}}

In the zero-photon subspace, the eigen-equation is given by $\hat{H}_{\text{sys}}\vert\Psi_{0,k}\rangle=\varepsilon_{0,k}\vert\Psi_{0,k}\rangle$ ($k=0,1,2,\cdots$) with the eigenstate $\vert\Psi_{0,k}\rangle=\vert0,0\rangle_{LR}\vert \tilde{k}(0,0)\rangle_{b}=\vert0,0\rangle_{LR}\vert k\rangle_{b}$ and the eigenvalue $\varepsilon_{0,k}=E_{0,0,k}$.

In the single-photon subspace, the eigensystem can be expressed as $\hat{H}^{[1]}_{\text{sys}}\vert\Psi_{1\pm,k}\rangle=\varepsilon_{1\pm,k}\vert\Psi_{1\pm,k}\rangle$ with
\begin{equation}
\hat{H}^{[1]}_{\text{sys}}=\left(
\begin{array}{ccccc}
E_{1,0,0} & J & 0 & 0 & \cdots  \\
J & E_{0,1,0} & 0 & 0 & \cdots  \\
0 & 0 & E_{1,0,1} & J & \cdots  \\
0 & 0 & J & E_{0,1,1} & \cdots  \\
\vdots  & \vdots  & \vdots  & \vdots  & \ddots
\end{array}
\right),
\end{equation}
which is written based on the basis states $\vert 1,0\rangle_{LR}\vert\tilde{k}(1,0)\rangle_{b}=(0,\cdots,0,1_{2k+1},0,\cdots)^{T}$ and $\vert 0,1\rangle_{LR}\vert\tilde{k}(0,1)\rangle_{b}=(0,\cdots,0,1_{2k+2},0,\cdots)^{T}$, where ``$T$'' denotes the matrix transpose. By solving the eigensystem of the matrix $\hat{H}^{[1]}_{\text{sys}}$, the eigenvalues and the eigenstates can be obtained as
\begin{equation}
\varepsilon_{1\pm,k}=\frac{1}{2}\left[E_{1,0,k}+E_{0,1,k}\pm\sqrt{(E_{1,0,k}-E_{0,1,k})^{2}+4J^{2}}\right],\qquad
\end{equation}
and
\begin{eqnarray}\label{Psi1}
\vert\Psi_{1\pm,k}\rangle&=&C_{1,0,k}^{[1\pm]}\vert1,0\rangle_{LR}\vert \tilde{k}(1,0)\rangle_{b}+C_{0,1,k}^{[1\pm]}\vert 0,1\rangle_{LR}\vert \tilde{k}(0,1)\rangle_{b}.\nonumber\\
\end{eqnarray}
The coefficients in Eq.~(\ref{Psi1}) are defined by
\begin{eqnarray}
C_{1,0,k}^{[1\pm]}&=&\frac{J}{\sqrt{J^{2}+(\varepsilon_{1\pm,k}- E_{1,0,k})^{2}}},\nonumber \\
C_{0,1,k}^{[1\pm]}&=&\frac{\varepsilon_{1\pm,k}-E_{1,0,k}}{\sqrt{J^{2}+(\varepsilon_{1\pm,k}-E_{1,0,k})^{2}}}.
\end{eqnarray}

\subsection{Eigensystem in the two-photon subspace \label{eigen2photon}}

In the two-photon subspace, the eigenstates and eigenvalues can be obtained by solving the eigen-equation $\hat{H}^{[2]}_{\text{sys}}\vert\Psi_{2s,k}\rangle =\varepsilon_{2s,k}\vert\Psi_{2s,k}\rangle$ with $s=\pm$, $0$ and $k=0,1,2,\cdots$. The matrix of $\hat{H}^{[2]}_{\text{sys}}$ can be expressed as
\begin{equation}
\hat{H}^{[2]}_{\text{sys}}=\left(
\begin{array}{ccccccc}
E_{2,0,0} & \sqrt{2}J & 0 & 0 & 0 & 0 & \cdots  \\
\sqrt{2}J & E_{1,1,0} & \sqrt{2}J & 0 & 0 & 0 &\cdots  \\
0 & \sqrt{2}J & E_{0,2,0} & 0 & 0 & 0 & \cdots  \\
0 & 0 & 0 & E_{2,0,1} & \sqrt{2}J & 0 & \cdots  \\
0 & 0 & 0 & \sqrt{2}J & E_{1,1,1} & \sqrt{2}J & \cdots  \\
0 & 0 & 0 & 0 & \sqrt{2}J & E_{0,2,1} & \cdots  \\
\vdots  & \vdots  & \vdots  & \vdots  & \vdots  & \vdots & \ddots
\end{array}
\right),
\end{equation}
which is defined based on the basis states $\vert 2,0\rangle_{LR}\vert \tilde{k}(2,0)\rangle_{b}=(0,\cdots,0,1_{3k+1},0,\cdots)^{T}$, $\vert 1,1\rangle_{LR}\vert \tilde{k}(1,1)\rangle_{b}=(0,\cdots,0,1_{3k+2},0,\cdots)^{T}$, and $\vert 0,2\rangle_{LR}\vert \tilde{k}(0,2)\rangle_{b}=(0,\cdots,0,1_{3k+3},0,\cdots)^{T}$. The eigenvalues can be obtained as
\begin{align}
\varepsilon_{2-,k}=&-\frac{1}{3}p_{k}-\frac{\sqrt{-3a_{k}}}{3}[\cos(\phi_{k}/3)+\sqrt{3}\sin(\phi_{k}/3)],  \nonumber \\
\varepsilon_{20,k}=&-\frac{1}{3}p_{k}-\frac{\sqrt{-3a_{k}}}{3}[\cos(\phi_{k}/3)-\sqrt{3}\sin(\phi_{k}/3)],  \nonumber \\
\varepsilon_{2+,k}=&-\frac{1}{3}p_{k}+\frac{2\sqrt{-3a_{k}}}{3}\cos(\phi_{k}/3),
\end{align}
where the relating parameters are defined by
\begin{align}
\phi_{k}&=\arccos [-3b_{k}\sqrt{-3a_{k}}/(2a_{k}^{2})],\nonumber \\
a_{k}&=q_{k}-p_{k}^{2}/3, \nonumber \\
b_{k}&=r_{k}+2p_{k}^{3}/27-p_{k}q_{k}/3,\nonumber \\
p_{k}&=-(E_{0,2,k}+E_{1,1,k}+E_{2,0,k}),\nonumber \\
q_{k}&=E_{1,1,k}E_{0,2,k}+E_{1,1,k}E_{2,0,k}+E_{2,0,k}E_{0,2,k}-4J^{2},\nonumber \\
r_{k}&=2J^{2}(E_{0,2,k}+E_{2,0,k})-E_{1,1,k}E_{0,2,k}E_{2,0,k}.
\end{align}
The corresponding eigenstates are
\begin{align}
\vert\Psi_{2s,k}\rangle&=C_{2,0,k}^{[2s]}\vert2,0\rangle_{LR}\vert\tilde{k}(2,0)\rangle_{b}\nonumber\\
&\quad+C_{1,1,k}^{[2s]}\vert 1,1\rangle_{LR}\vert\tilde{k}(1,1)\rangle_{b}\nonumber\\
&\quad+C_{0,2,k}^{[2s]}\vert 0,2\rangle_{LR}\vert\tilde{k}(0,2)\rangle_{b},
\end{align}
where the superposition coefficients are given by
\begin{align}
C_{2,0,k}^{[2s]} &=\sqrt{2}J(E_{0,2,k}-\varepsilon_{2s,k})N_{2s,k}^{-1/2},\nonumber \\
C_{1,1,k}^{[2s]} &=(E_{2,0,k}-\varepsilon_{2s,k})(E_{0,2,k}-\varepsilon_{2s,k})N_{2s,k}^{-1/2},\nonumber \\
C_{0,2,k}^{[2s]} &=\sqrt{2}J(E_{2,0,k}-\varepsilon_{2s,k})N_{2s,k}^{-1/2},
\end{align}
with
\begin{align}
N_{2s,k}&=[2J^{2}+(E_{2,0,k}-\varepsilon_{2s,k})^{2}](E_{0,2,k}-\varepsilon_{2s,k})^{2}\nonumber \\
&\quad+2J^{2}(E_{2,0,k}-\varepsilon_{2s,k})^{2}.
\end{align}

\begin{widetext}
\section{Derivation of the probability amplitudes \label{appendixb}}
In this Appendix, we present the detailed derivation of the probability amplitudes. Based on the Schr\"{o}dinger equation $i\vert\dot{\psi}(t)\rangle=\hat{H}_{\text{eff}}\vert\psi(t)\rangle$ with $\hat{H}_{\text{eff}}$ and $\vert\psi(t)\rangle$ defined by Eqs.~(\ref{Heff}) and (\ref{psit}), the equations of motion of these probability amplitudes $C_{m,n,k}(t)$ can be obtained as
\begin{subequations}
\label{equproamp}
\begin{align}
i\dot{C}_{0,0,k}(t)=&\ C_{0,0,k}(t)E_{0,0,k}+\Omega\sum_{l=0}^{\infty}C_{1,0,l}(t)_{b}\!\langle \tilde{k}(0,0)\vert\tilde{l}(1,0)\rangle_{b},  \\
i\dot{C}_{0,1,k}(t)=&\ C_{0,1,k}(t)\left(E_{0,1,k}-i\frac{\kappa_{R}}{2}\right)+JC_{1,0,k}(t)+\Omega\sum_{l=0}^{\infty}C_{1,1,l}(t)_{b}\!\langle \tilde{k}
(0,1)\vert\tilde{l}(1,1)\rangle_{b},   \\
i\dot{C}_{1,0,k}(t)=&\ C_{1,0,k}(t)\left(E_{1,0,k}-i\frac{\kappa_{L}}{2}\right)+JC_{0,1,k}(t)+\Omega\sum_{l=0}^{\infty}[C_{0,0,l}(t)_{b}\!\langle \tilde{k}(1,0)\vert\tilde{l}(0,0)\rangle_{b}+\sqrt{2}C_{2,0,l}(t)_{b}\!\langle\tilde{k}(1,0)\vert\tilde{l}(2,0)\rangle_{b}],   \\
i\dot{C}_{0,2,k}(t)=&\ C_{0,2,k}(t)(E_{0,2,k}-i\kappa_{R})+J\sqrt{2}C_{1,1,k}(t),   \\
i\dot{C}_{2,0,k}(t)=&\ C_{2,0,k}(t)(E_{2,0,k}-i\kappa_{L})+J\sqrt{2}C_{1,1,k}(t)+\sqrt{2}\Omega\sum_{l=0}^{\infty}C_{1,0,l}(t)_{b}\!\langle \tilde{k}(2,0)\vert \tilde{l}(1,0)\rangle_{b},   \\
i\dot{C}_{1,1,k}(t)=&\ C_{1,1,k}(t)\left[E_{1,1,k}-i\left(\frac{\kappa_{L}}{2}+\frac{\kappa_{R}}{2}\right)\right]+\sqrt{2}JC_{0,2,k}(t)+\sqrt{2}JC_{2,0,k}(t)+\Omega \sum_{l=0}^{\infty}C_{0,1,l}(t)_{b}\!\langle\tilde{k}(1,1)\vert\tilde{l}(0,1)\rangle_{b},
\end{align}
\end{subequations}
where the eigenvalues $E_{m,n,k}$ are given by Eq.~(\ref{eigenvalues}) and the photon-number-dependent displaced phonon number states $\vert\tilde{k}(m,n)\rangle_{b}$ are given by Eq.~(\ref{disphononnum}).

Under the weak-driving condition ($\Omega\ll\kappa_{L}$), the equations of motion of these probability amplitudes $C_{m,n,k}(t)$ can be approximately solved by using a perturbation method, i.e., discarding the higher-order terms in the equations of motion for the lower-order variables. Hence, the equations of motion of probability amplitudes become
\begin{subequations}
\label{equproampper}
\begin{align}
i\dot{C}_{0,0,k}(t)\approx&\ C_{0,0,k}(t)E_{0,0,k},  \\
i\dot{C}_{0,1,k}(t)\approx&\ C_{0,1,k}(t)\left(E_{0,1,k}-i\frac{\kappa}{2}\right)+JC_{1,0,k}(t),   \\
i\dot{C}_{1,0,k}(t)\approx&\ C_{1,0,k}(t)\left(E_{1,0,k}-i\frac{\kappa}{2}\right)+JC_{0,1,k}(t)+\Omega\sum_{l=0}^{\infty}C_{0,0,l}(t)_{b}\!\langle \tilde{k}%
(1,0)\vert\tilde{l}(0,0)\rangle_{b},   \\
i\dot{C}_{0,2,k}(t)=&\ C_{0,2,k}(t)(E_{0,2,k}-i\kappa)+J\sqrt{2}C_{1,1,k}(t),   \\
i\dot{C}_{2,0,k}(t)=&\ C_{2,0,k}(t)(E_{2,0,k}-i\kappa)+J\sqrt{2}C_{1,1,k}(t)+\sqrt{2}\Omega\sum_{l=0}^{\infty}C_{1,0,l}(t)_{b}\!\langle \tilde{k}(2,0)\vert \tilde{l}(1,0)\rangle_{b},   \\
i\dot{C}_{1,1,k}(t)=&\ C_{1,1,k}(t)(E_{1,1,k}-i\kappa)+\sqrt{2}JC_{0,2,k}(t)+\sqrt{2}JC_{2,0,k}(t)+\Omega \sum_{l=0}^{\infty}C_{0,1,l}(t)_{b}\!\langle\tilde{k}(1,1)\vert\tilde{l}(0,1)\rangle_{b},
\end{align}
\end{subequations}
where we considered the case of $\kappa_{L}=\kappa_{R}=\kappa$. In the case of $g_{L}=g_{R}=g$, the matrix elements in Eq.~(\ref{equproampper}) satisfy
\begin{equation}
_{b}\!\langle \tilde{k}(1,0)\vert\tilde{l}(0,0)\rangle_{b}=\,_{b}\!\langle \tilde{k}(2,0)\vert \tilde{l}(1,0)\rangle_{b}
=\,_{b}\!\langle\tilde{k}(1,1)\vert\tilde{l}(0,1)\rangle_{b}=\,_{b}\!\langle k\vert D(-g/\omega_{M})\vert l\rangle_{b}\equiv\Pi_{k,l}.
\end{equation}
We assume $C_{0,0,0}=1$, then the steady-state solutions of Eq.~(\ref{equproampper}) can be approximately obtained by setting $\dot{C}_{m,n,k}(t)=0$ as
\begin{subequations}
\label{proampper}
\begin{align}
C_{0,0,k\neq0}=&\ 0,  \\
C_{0,1,k}=&\ -\frac{4J\Omega\Pi_{k,0}}{4J^{2}-4E_{0,1,k}E_{1,0,k}+2i(E_{0,1,k}+E_{1,0,k})\kappa+\kappa^{2}},  \\
C_{1,0,k}=&\ \frac{2\Omega(2E_{0,1,k}-i\kappa)\Pi_{k,0}}{4J^{2}-4E_{0,1,k}E_{1,0,k}+2i(E_{0,1,k}+E_{1,0,k})\kappa+\kappa^{2}},   \\
C_{0,2,k}=&\ \frac{\sqrt{2}J\Omega\left[2J\sum_{l=0}^{\infty}C_{1,0,l}\Pi_{k,l}+(i\kappa-E_{2,0,k})\sum_{l=0}^{\infty}C_{0,1,l}\Pi_{k,l}\right]}
{2J^{2}(E_{2,0,k}-i\kappa)+(E_{0,2,k}-i\kappa)[2J^{2}+(iE_{1,1,k}+\kappa)(iE_{2,0,k}+\kappa)]},   \\
C_{2,0,k}=&\ -\frac{\sqrt{2}\Omega\{J(E_{0,2,k}-i\kappa)\sum_{l=0}^{\infty}C_{0,1,l}\Pi_{k,l}+[2J^{2}+(iE_{1,1,k}+\kappa)(iE_{0,2,k}+\kappa
)]\sum_{l=0}^{\infty}C_{1,0,l}\Pi_{k,l}\}}{2J^{2}(E_{0,2,k}+E_{2,0,k})-E_{0,2,k}E_{1,1,k}E_{2,0,k}+i(M_{k}-4J^{2})\kappa
+(E_{0,2,k}+E_{1,1,k}+E_{2,0,k})\kappa^{2}-i\kappa^{3}},   \\
C_{1,1,k}=&\ \frac{\Omega(E_{0,2,k}-i\kappa)[(E_{2,0,k}-i\kappa)\sum_{l=0}^{\infty}C_{0,1,l}\Pi_{k,l}
-2J\sum_{l=0}^{\infty}C_{1,0,l}\Pi_{k,l}]}{2J^{2}(E_{0,2,k}+E_{2,0,k})-E_{0,2,k}E_{1,1,k}E_{2,0,k}+i(M_{k}-4J^{2})\kappa
+(E_{0,2,k}+E_{1,1,k}+E_{2,0,k})\kappa^{2}-i\kappa^{3}},
\end{align}
\end{subequations}
where we introduce the coefficient
\begin{equation}
M_{k}=E_{1,1,k}E_{2,0,k}+E_{0,2,k}E_{1,1,k}+E_{0,2,k}E_{2,0,k}.
\end{equation}

\end{widetext}

\end{document}